\newcommand{\comment}[1]{}

\documentclass[12pt]{article}

\usepackage[english]{babel}

\usepackage[letterpaper,top=2cm,bottom=2cm,left=3cm,right=3cm,marginparwidth=1.75cm]{geometry}

\usepackage{amsmath}
\usepackage{graphicx}
\usepackage[colorlinks=true, allcolors=black]{hyperref}

\usepackage{authblk}
\usepackage{float}
\usepackage{siunitx}

\usepackage{chemformula}

\usepackage{times}

\title{Tunneling in a very slow ion-molecule reaction}

\author{Robert Wild}
\author{Markus N{\"o}tzold}
\author{Malcolm Simpson}
\author{Thuy Dung Tran\footnote{Present address: Department of Optics, Palack{\'y} University, 17. listopadu 12, 77146 Olomouc, Czech Republic}}
\author{Roland Wester\footnote{To whom correspondence should be addressed; E-mail: roland.wester@uibk.ac.at}}

\affil{Institut f{\"u}r Ionenphysik und Angewandte Physik, Universit{\"a}t Innsbruck, Technikerstraße 25, 6020 Innsbruck, Austria}

\date{}

\begin{document} 




\maketitle

\textbf{Quantum tunneling reactions play a significant role in chemistry when classical pathways are energetically forbidden \cite{McMahon2003:s}, be it in gas phase reactions, surface \textcolor{black}{diffusion}, or liquid phase chemistry. \textcolor{black}{In general, such tunneling reactions are challenging to calculate theoretically, given the high dimensionality of the quantum dynamics, and also very difficult to identify experimentally \cite{Shannon2013:nc,Tizniti2014:nc,Yang2019:nc}.} Hydrogenic systems, however, allow for accurate first-principles calculations. In this way the rate of the \textcolor{black}{gas phase} proton transfer tunneling reaction of hydrogen molecules with deuterium anions, H$_2$ + D$^-$ $\rightarrow$ H$^-$ + HD, has been calculated \cite{Yuen2018:pra}, but has so far lacked experimental verification. Here we present high-sensitivity measurements of the reaction rate carried out in a cryogenic 22-pole ion trap. We observe an extremely low rate constant of $\mathbf{(5.2\pm1.6) \times 10^{-20}}$ $\mathbf{\mathrm{cm}^{3}/s}$. This measured value agrees with quantum tunneling calculations, serving as a benchmark for molecular theory and advancing the understanding of fundamental collision processes. A deviation of the reaction rate from linear scaling, which is observed at high H$_2$ densities, can be traced back to previously unobserved heating dynamics in radiofrequency ion traps.}

Hydrogen is the most abundant element in the universe \cite{Ferriere2001:rmp}, and collisions of hydrogen and its charged forms are important in the chemistry and evolution of the interstellar medium \cite{Gerlich2002:pss,Tielens2013:rmp}. Binary collisions of atomic with molecular hydrogen belong to the most fundamental molecular systems to be studied experimentally and are simple enough to be investigated using full quantum calculations \cite{Kitsopoulos1993:s,Harich2002:n}. Here we investigate the presumably most fundamental ion-molecule reaction, the proton transfer reaction from hydrogen molecules to hydrogen atomic anions. This reaction, which can be made exoergic by virtue of vibrational zero-point energy differences when working with deuterium anions, may proceed via tunneling through its intermediate barrier (see Fig.\ 1a),
\begin{equation}
\mathrm{D^- + H_2 \rightarrow H^- + HD}.
\end{equation}
The collision complex has the makeup of the H$_3^-$ anion. \textcolor{black}{Its stability and linear structure has already been proposed in 1937 \cite{Stevenson1937:jcp}, but later questioned due to the difficulty of accurately calculating the vibrational zero point energy. Following the calculation of the first three-dimensional potential energy surface its stability could be theoretically confirmed \cite{Staerck1993:cp} and later verified experimentally \cite{Wang2003:cpl}.} Interstellar H$_3^-$ could function as a tracer for H$^-$, which is thought to be abundantly present in the interstellar medium but not yet observed, mainly because it has only one bound electronic state \cite{Ayouz2011:pra}.

\textcolor{black}{Experimental investigations of reaction (1) at energies of a few electronvolts and below started in the 1990s. Crossed-beam experiments revealed state-specific cross sections and angular distributions for an isotopic variant of reaction (1) \cite{Zimmer1995:jpb}. These were compared to pioneering quantum scattering calculations \cite{Belyaev1993:cpl}. In a guided ion beam experiment, the barrier height of reaction (1) was found to be about 330 meV \cite{Haufler1997:jpca}.} \textcolor{black}{Numerous further} theoretical investigations followed, most of which considered energies above the potential barrier
(see e.\ g.\ \cite{Giri2006:jpca,Zhang2010:cpl,Wang2013:jpca}). For temperatures in the range of a few Kelvin, colliding reactants can generally not overcome this barrier but will chiefly react via tunneling, the probability of which typically increases with lower temperature because of greater lifetimes of intermediate ion-molecule complexes \cite{Mikosch2008:jpca}. In ion-molecule collisions, only two tunneling reactions have been discovered so far, NH$_3^+$ + H$_2$ and c-C$_3$H$_2^+$ + H$_2$ \cite{Herbst1991:jcp,Markus2020:prl}. A semiclassical estimate of the tunneling probability of reaction (1) suggested a rate coefficient of about $10^{-19}$\,cm$^3$/s \cite{Endres2017:pra}. In a previous measurement only an upper bound could be provided \cite{Endres2017:pra}. This result prompted ab initio quantum calculations of the reaction probability at low collision energies, and predicted a rate constant two orders of magnitude lower than the experimental upper limits \cite{Yuen2018:pra}.

Here we present a measurement of the low temperature reaction rate constant of equation (1), providing a benchmark for molecular quantum tunneling theory. The principle of this measurement is straightforward: negatively charged deuterium ions are loaded into a 22-pole linear radiofrequency (RF) trap, mounted on a closed-cycle helium cryostat \cite{Gerlich1995:ps}. The 22-pole configuration provides a large field-free region to minimize radiofrequency heating, as sketched in Fig.\ 1a \cite{Wester2009:jpb,Asvany2009:ijms}. The D$^-$ ions are collisionally cooled with 10\,K buffer gas of hydrogen, which leads to an estimated collision temperature of $15\pm5$\,K due to radiofrequency heating (see methods section). Once the ions are cooled, additional hydrogen gas is added to the trap and maintained at constant density. After a chosen interaction time the ions are ejected from the trap and mass-selectively measured via time-of-flight mass spectrometry.

To measure typical ion-molecule reaction rate coefficients, which have magnitudes around 10$^{-9}$\,cm$^3$/s, neutral densities of the order of (10$^{10}$-10$^{12}$)/cm$^3$ are commonly used. This implies that trapping times of 10\,ms to 1\,s are sufficient to observe the reaction kinetics. However, to study rate coefficients smaller than 10$^{-18}$\,cm$^3$/s, densities beyond 10$^{15}$/cm$^3$ and at the same time trapping times up to 1000\,s are required (see methods section). Fig.\ 1b shows the time-dependent appearance of a small peak of H$^-$ ions after 950\,s of trapping time at a density of $2.8\times10^{14}$/cm$^3$, which signifies that reaction (1) is actually taking place. Despite the high densities, no evidence for products from three-body recombination are detected.

Fig.\ 2 shows examples for measurements of the H$^-$ fraction as a function of interaction time for different H$_2$ densities. We fit this fraction to the solution of rate equations (Eq. 4--5) which include the growth of H$^-$ and a background loss rate of H$^-$ and D$^-$:
\begin{equation}\label{eqn:rate2}
	\frac{N_\mathrm{H}(t)}{N_\mathrm{D}(0)}
    =
    \left(\frac{k^\mathrm{gr}_{\mathrm{H}}}{k^\mathrm{bg}_{\mathrm{H}}-k^\mathrm{bg}_{\mathrm{D}}-k^\mathrm{gr}_{\mathrm{H}}}\right)\left(\mathrm{e}^{-(k^\mathrm{gr}_{\mathrm{H}}+k^\mathrm{bg}_{\mathrm{D}})t}-\mathrm{e}^{-k^\mathrm{bg}_{\mathrm{H}}t}\right)
    +
    \frac{N_\mathrm{H}(0)}{N_\mathrm{D}(0)}\mathrm{e}^{-k^\mathrm{bg}_{\mathrm{H}}t},
\end{equation}
where $k^\mathrm{gr}_{\mathrm{H}}$ is the growth rate of H$^-$. $k^\mathrm{bg}_{\mathrm{X}}$ denotes the respective background loss rates of H$^-$ and D$^-$ out of the trap, and $N_\mathrm{H}(0)$ and $N_\mathrm{D}(0)$ are the initial amounts of H$^-$ and D$^-$. A separate fit to the decay of the D$^-$ peak provides the sum of the D$^-$ background loss rate and the H$^-$ growth rate. The small amount of initial H$^-$ comes from the trap loading: as high-energy D$^-$ collides with the H$_2$ buffer gas during initial cooling, some H$^-$ is created via classical over-the-barrier reactions.

The H$^-$ growth rates that are extracted from the fits are plotted as a function of H$_2$ density in Fig.\ 3a; the background loss rates are shown in Fig.\ 3b. At the three lowest densities, the growth rates are consistent with a \textcolor{black}{fit to} linear dependence $k^\mathrm{gr}_{\mathrm{H}}=k_\mathrm{r} n$ with $k_\mathrm{r}=(5.2\pm1.6) \times 10^{-20}$ $\mathrm{cm}^{3}$/s \textcolor{black}{(black dashed line in Fig.\ 3a)}. This rate coefficient corresponds roughly to one reaction occurring for every $10^{11}$ collisions, \textcolor{black}{when the rate of all collisions that overcome the centrifugal barrier is described by the Langevin capture rate coefficient, which is about $2\times 10^{-9}$ $\mathrm{cm}^{3}$/s.} This is to our knowledge the lowest measured bimolecular ion-molecule reaction rate constant by four orders of magnitude, see e.\ g.\ the Kinetic Database for Astrochemisty (KIDA) \cite{Wakelam2012:ajss}. The rate coefficient agrees very well with the \textcolor{black}{quantum} theoretical prediction for the tunneling rate coefficient for normal-H$_2$ colliding with a thermal D$^-$ velocity distribution at a temperature of 15\,K \cite{Yuen2018:pra}, which is indicated by the \textcolor{black}{violet} line in Fig.\ 3a. The width of the line represents the finite uncertainty of the ion temperature. \textcolor{black}{A previous semiclassical statistical calculation based on an earlier potential energy surface provided about three orders of magnitude lower tunneling rate coeffficient, which shows the enormous sensitivity of tunneling rates on the theoretical methods \cite{Luo2011:ctc}.}

At higher densities a strong deviation of the reaction rate from linear behavior appears, accompanied by an increase of trap loss rate of H$^-$, as seen in Fig.\ 3b. Such a nonlinear behavior has not been reported before for ion-molecule kinetics in traps. To understand this phenomenon, one must take a closer look at the velocity distributions in radiofrequency ion traps. Using the adiabatic approximation, the ion velocity can be separated into micromotion, which is caused by the RF voltages and gives rise to the effective trapping potential, and secular motion, describing classical trajectories within the effective potential \cite{Gerlich1995:ps,Wester2009:jpb}. If an ion undergoes a collision during micromotion, energy can be transferred from the RF fields to the secular motion, which is known as collisionally induced RF heating and has plagued ion traps since their invention. Due to this RF heating, the velocity distribution is no longer described well by Maxwell-Boltzmann statistics but brings about a high-energy tail in the ion distribution \cite{Asvany2009:ijms}. This must in general be considered for kinetics measurements, as a small number of reactions from the high-energy tail could skew the average measured reaction rates.

Following the nonextensive generalization of entropy by Tsallis \cite{Tsallis1988:jsp}, the velocity distribution in one dimension $f(v_i)$ of a trapped ion of mass $m$ subjected to RF heating can be described by a $q$-exponential Tsallis distribution,
\begin{equation}
	f(v_i)\propto\left( 1+(q-1)\frac{m v_i^2}{2k_BT}\right)^{\frac{1}{1-q}}
\end{equation}
which in addition to the temperature $T$ contains the parameter $q$ that describes the strength of a power-law tail and can be generalized to the full velocity distribution \textcolor{black}{(see Eqs. 6--9 in the methods section)}. The limit $q\rightarrow1$ recovers the Gaussian probability curve of a Maxwellian gas. For collision dynamics the distribution of relative velocities between ions and neutrals is the relevant quantity. We obtain these by simulating the relative velocities between the Tsallis distribution of the ions and the Maxwell-Boltzmann distribution of the buffer gas.

The \textcolor{black}{simulated relative velocity distributions are} plotted in Fig.\ 3c for two different hydrogen densities and fitted with Tsallis distribution functions. The high-energy tail is notably increased at the high density, a phenomenon not observed in multipole ion traps before. This can be qualitatively understood to stem from a small mean-free-path of the trapped ions in the dense buffer gas. In multipole ion traps RF heating occurs close to the turning points, where the micromotion amplitude is large. At standard buffer gas densities ions make multiple round trips through the trap between buffer gas collisions. Thus the probability for a subsequent collision to cool rather than heat is high. Then the ion velocity distribution has suppressed high-energy tails and is density independent. However, at densities above $10^{15}$ $\mathrm{cm}^{-3}$, the mean-free-path becomes smaller than the trap diameter and comparable to the distances over which radiofrequency heating occurs (see Extended Data Fig.\ 1). This can reduce the ion temperature in the trap center but causes a ``run-away'' heating of the ions near the trap edges, which manifests itself as increased high-energy tails.

To determine the influence of the increased energy tails on the D$^-$ reaction rate, we integrate the density-dependent Tsallis distributions for the relative velocity over the theoretical reaction probability \textcolor{black}{(see Fig.\ 6 of Ref. \cite{Yuen2018:pra})} \textcolor{black}{(see methods section for details)}. The resulting reaction rate is plotted as open circles in Fig.\ 3a\textcolor{black}{, the error bars represent the standard deviation of several simulations.} This simulation shows an enhanced rate at high densities in good agreement with the nonlinear increase of the measured rate. Only the magnitude of the rate increase is slightly lower compared to the measurements. We expect that this extra heating is caused by differences between the real and simulated electric fields due to trap imperfections and patch potentials \textcolor{black}{(see Extended Data Fig.\ 2)}.

In the low density limit run-away heating is not present. The rate measurements for the lowest three densities in Fig.\ 3a agree well with a linear density dependence, also the H$^-$ loss rates remain fairly constant here. Thus the reaction rate coefficient can be determined from the slope of these data points, as stated above and plotted in Fig.\ 4. Also shown in this figure is the theoretical tunneling rate coefficient of Yuen {\it et al.} \cite{Yuen2018:pra}. The black and red \textcolor{black}{long-dashed} lines show the theory for para- and ortho-H$_2$, respectively, with the \textcolor{black}{violet solid} line being the weighted average and representing the normal-H$_2$ that is used in the present experiment. This shows again that the measured rate coefficient agrees very well with the theoretical rate for normal-H$_2$. In this temperature range the rate coefficient is dominated by tunneling. This becomes particularly clear from the \textcolor{black}{short-dashed} lines in Fig.\ 4, which show the theoretical rate coefficient when only over-the-barrier reactions without tunneling are considered.

In this work we have measured the reaction rate constant of the proton exchange reaction D$^-$ with H$_2$, which occurs by quantum mechanical tunneling through the reaction barrier. The extremely small measured value of $(5.2\pm1.6) \times 10^{-20}$\,cm$^3$/s is in very good agreement with ab initio quantum scattering calculations. High-energy tails were found to influence the kinetics at high densities and at temperatures where over-the-barrier reactions become active. This is an effect that will have to be included in other ion trap reaction measurements as well, in particular for endothermic or barrier-limited reactions. From our simulations we know that these high-energy tails play only a minor role at the lower densities where tunneling takes place. This work shows how not only spectroscopic investigations, but also reaction kinetics can benchmark accurate quantum theoretical calculations. As such this broadens the understanding of tunneling processes \textcolor{black}{and will open up further research on the quantum dynamics of neutral and ionic reactions, for example on the role of scattering resonances.}


\clearpage

\begin{figure}[p]
	\centering
	\includegraphics[width=\columnwidth]{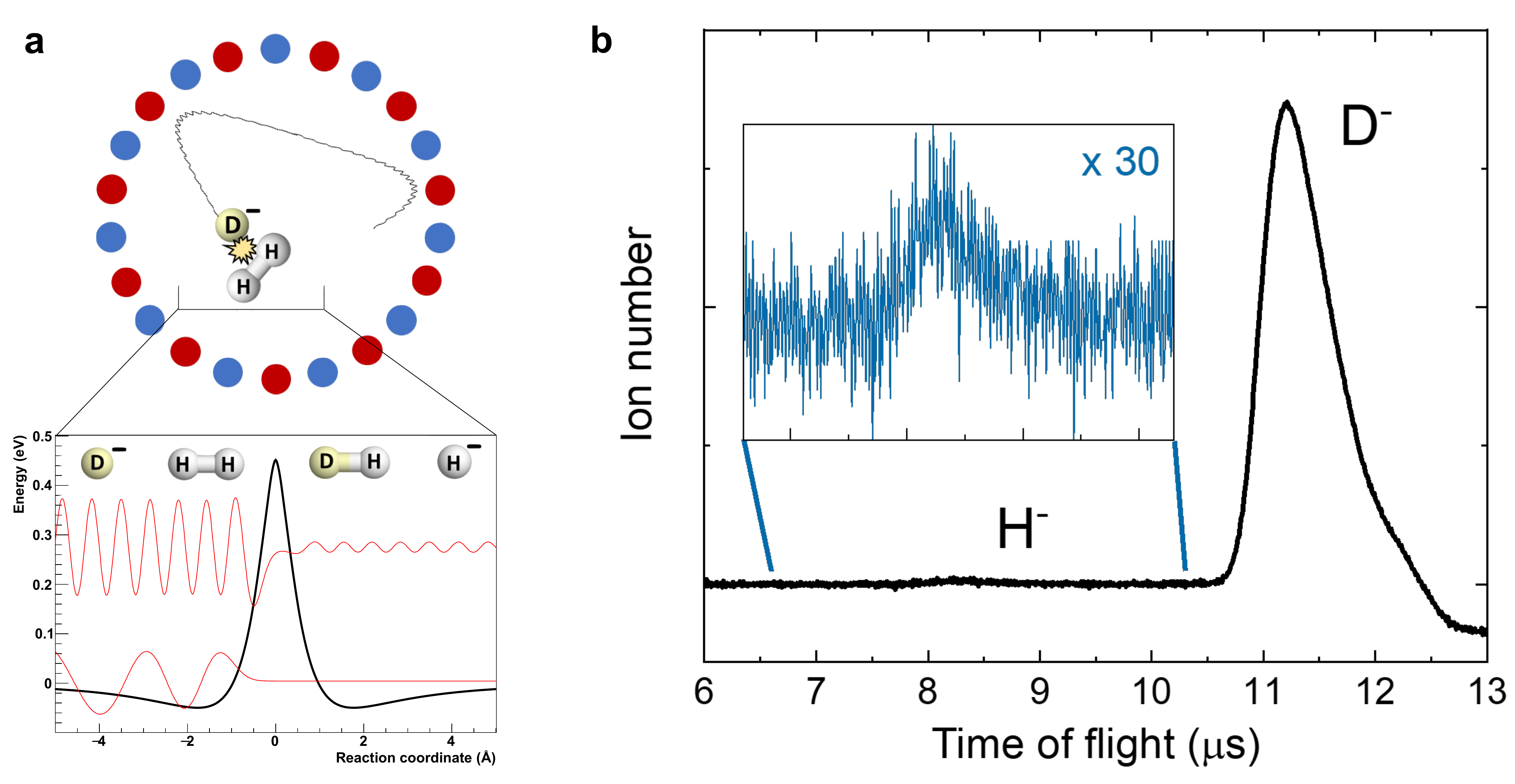} 
	\caption{{\bf Overview of the experiment.} {\bf a} Cross section of the 22-pole ion trap, viewed along the symmetry axis. An RF voltage is alternately applied to the trap rods (red and black), trapping D$^-$ ions. During collisions with H$_2$, proton exchange may occur via tunneling through the reaction barrier\textcolor{black}{, as indicated by the schematic wavefunctions crossing the mininmum energy path \cite{Endres2017:pra} of the reaction}. {\bf b}  The ion signal measured at the microchannel plate detector, after time-of-flight separation. Shown is the average of 16 time-of-flight traces, taken at the lowest H$_2$ density of $2.8\times10^{14}$ $\mathrm{cm}^{-3}$ and 950 seconds interaction time. The inset is the signal multiplied by 30 to emphasize the H$^-$ peak.}
\end{figure}

\clearpage

\begin{figure}
	\centering
	\includegraphics[width=\columnwidth]{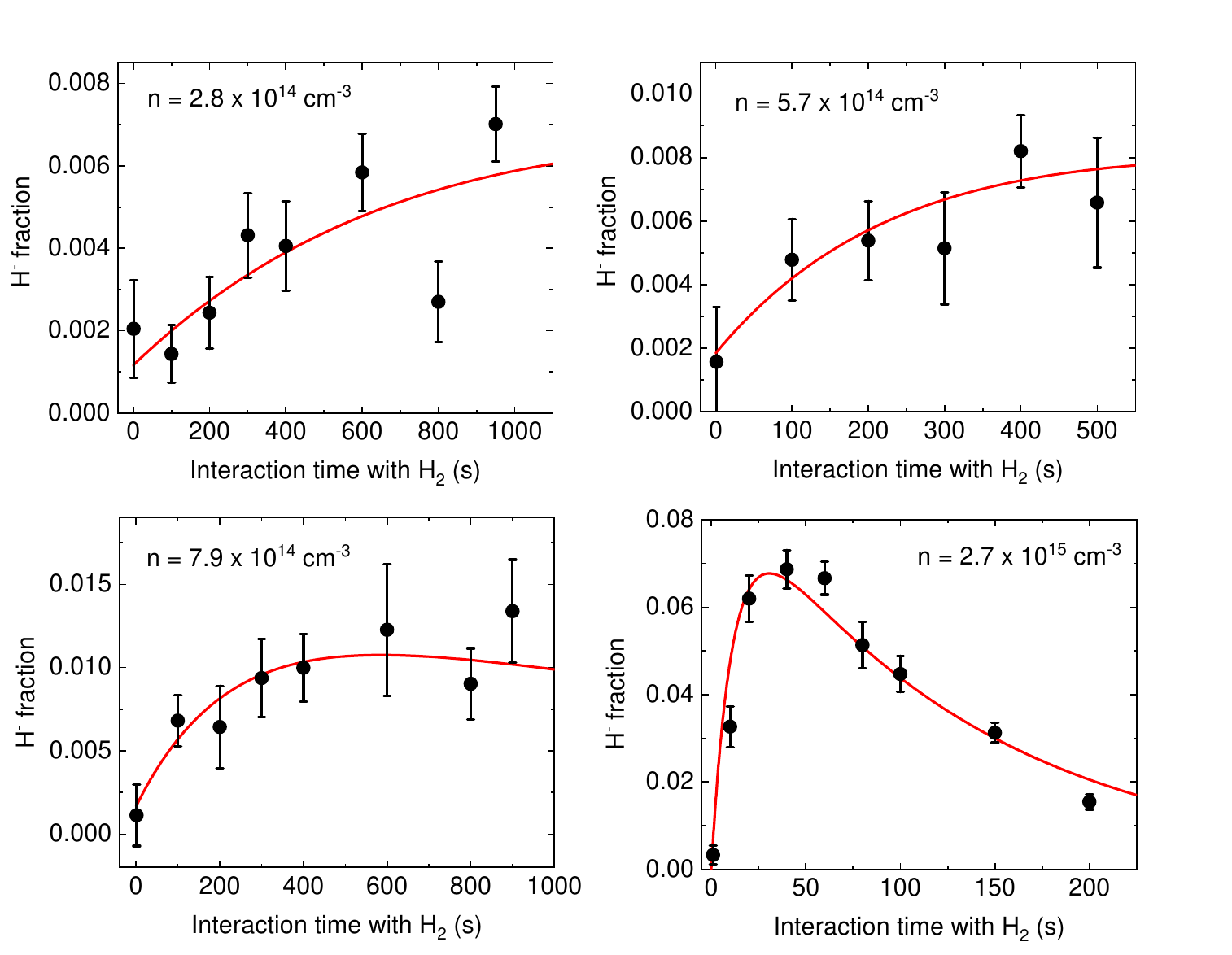}
	\caption{{\bf Ion-molecule reaction kinetics.} The fractional amount of H$^-$ present in the trap as a function of time is shown at four different densities. Red lines are fits to equation (2). As density increases, the loss of H$^-$ out of the trap becomes prominent.}
\end{figure}

\begin{figure}
	\centering
  \includegraphics[width=\columnwidth]{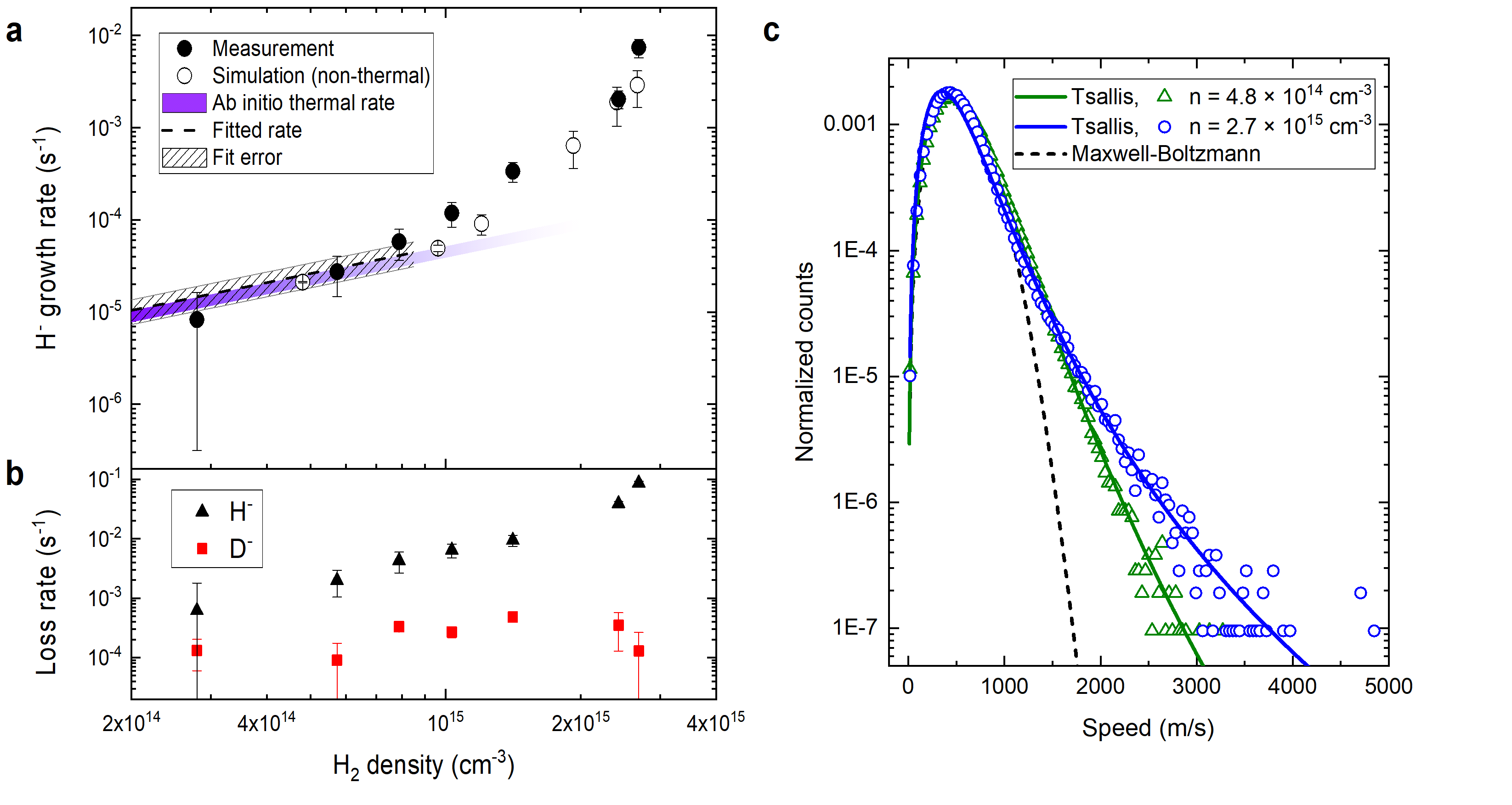}
  \caption{{\bf Density dependence of rates and velocity distributions.} {\bf a} Black points show measurements of the H$^-$ growth rate. The violet line is the theoretical prediction for the tunneling rate \textcolor{black}{for normal-H$_2$} at the estimated collision temperature of $15\pm5$\,K. The open circles show the result of a numerical simulation. \textcolor{black}{The black dashed line shows the linear fit to the measured rates at the lower densities.} {\bf b} Loss rates of the parent and product ions out of the trap. Elevated loss rates are seen above $7\times10^{14}$ $\mathrm{cm}^{-3}$ of H$_2$, which is where the H$^-$ growth rate begins to deviate from linear. {\bf c} The distribution of relative velocities between the ions and a 10\,K Maxwell-Boltzmann hydrogen gas for two different buffer gas densities. The distributions are fitted with a tail-weighted Tsallis function (equation (3)). A Maxwell-Boltzmann fit is shown for reference.}
\end{figure}

\begin{figure}
	\centering
	\includegraphics[width=\columnwidth]{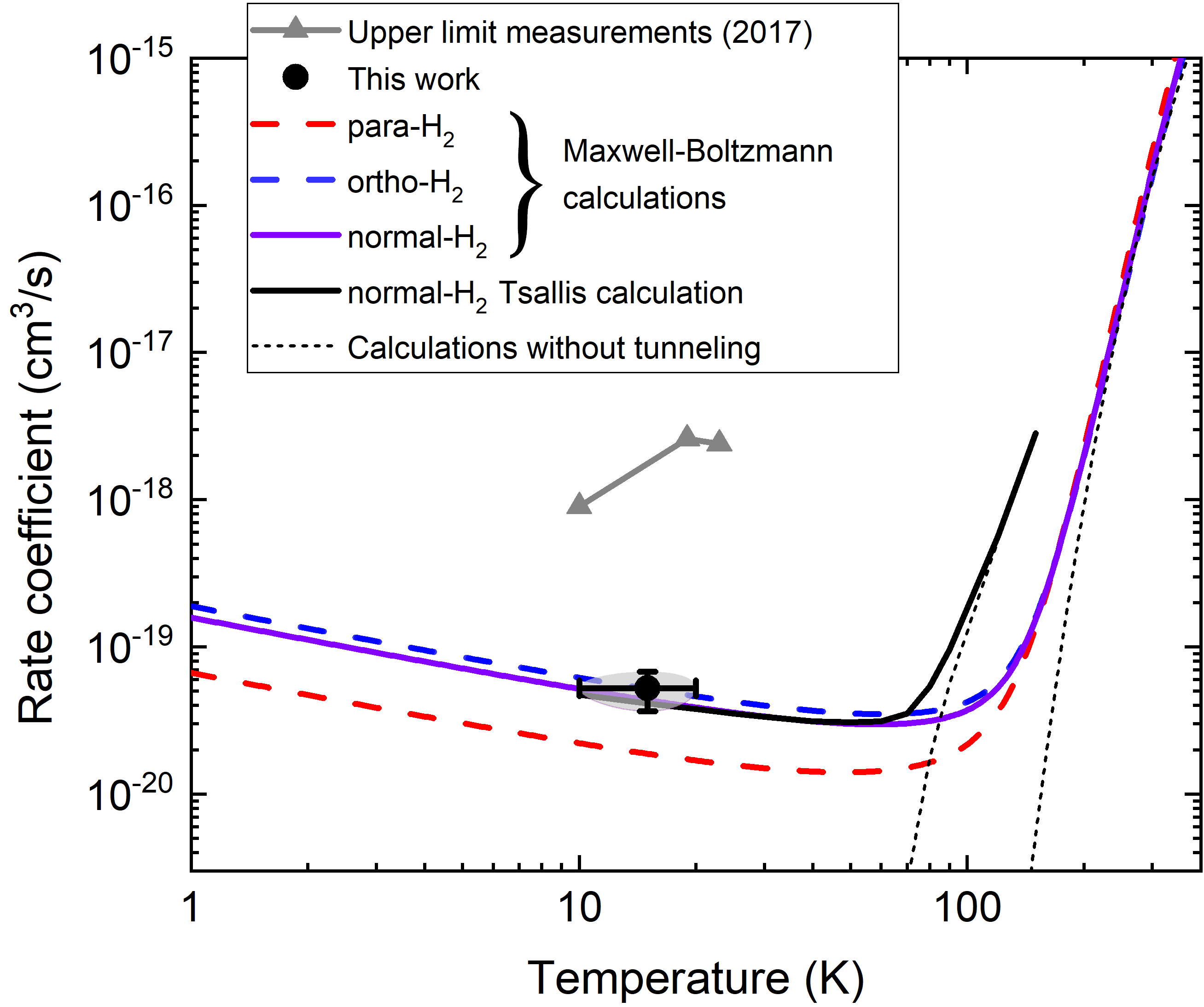}
	\caption{{\bf Tunneling rate coefficient in comparison with theory.} Measured and calculated reaction rate coefficients as a function of collision temperature. The black point with the error ellipse is the measurement presented in this work. Red and black \textcolor{black}{long-dashed} lines show the Maxwell-Boltzmann averages of the calculation for ortho- and para-H$_2$ \cite{Yuen2018:pra}, with the \textcolor{black}{violet solid} line the as weighted average \textcolor{black}{representing normal-H$_2$}. The solid black line shows the simulated result including the high-energy tail of the ion velocity distribution. The \textcolor{black}{short-dashed} black lines indicate the theoretical expectations (with and without high-energy tail) when tunneling is excluded. The grey triangles show the upper limit measurement of 2017 \cite{Endres2017:pra}.}
\end{figure}

\clearpage

\subsection*{Methods}

\subsubsection*{Experiment}
	
For the ion-molecule reaction experiments we use a 22-pole linear radiofrequency trap configuration which is described in detail elsewhere \cite{Gerlich1995:ps,Wester2009:jpb,Best2011:aj,Endres2017:jms}. We create D$^-$ and a small amount of H$^-$ in a plasma discharge of deuterium gas. A Wiley-McLaren spectrometer accelerates the ions toward the trapping region, and we selectively load D$^-$ into the multipole trap by time-of-flight (ToF) mass separation. Within the trap the ions collide with normal-H$_2$ buffer gas (75\% ortho and 25\% para H$_2$) that has thermalized with the trap's copper housing at 10\,K. \textcolor{black}{At such low temperatures, measurements of average ion temperatures repeatedly show temperatures warmer than the buffer gas \cite{Jusko2014:prl,Endres2017:jms} due to radiofrequency heating. Based on this, we estimate the collision temperature for the D$^-$ + H$_2$ collisions to be $15\pm5$\,K.} Both the imperfect ToF separation and collisions of D$^-$ with H$_2$ during the initial cooling contribute to some initial H$^-$ in the trap. Using helium as a buffer gas would avoid the latter, but we have found the mass difference beween He and D$^-$ to have an adverse effect on trap lifetimes.

To be sensitive to low reaction rates one desires high densities and long interaction times. Very high reactant gas densities can cause discharges at the high voltage of the microchannel plate (MCP) detector. To avoid this the MCP voltage is turned off before reactant gas is added. After the interaction time the gas flow is switched off and the chamber pumped down until the pressure at the MCP detector is below $10^{-6}$ mbar. To probe long interaction times, extra effort was made to keep the vacuum chamber free of stray reactants such as water molecules, which limited experiments in the past \cite{Endres2017:pra}. In the measurements presented here background lifetimes of D$^-$ were on the order of three hours.

Pressures of H$_2$ are measured with a capacitive gauge connected to the trap housing via a teflon tube. The capacitive gauge allows for an absolute pressure measurement independent of gas species. When calculating the gas densities in the trap, corrections for the temperature difference between the trap and the gauge, as well as thermal transpiration effects, were performed \cite{Endres2017:pra}. The effects of thermal transpiration in the pressure regime in which we operate are pressure corrections ranging from 25\% to 60\%. Based on these we estimate a 10\% error in absolute density.

During extraction from the trap, we pulse a small gate voltage to deflectors situated close to the trap opening to allow only a small fraction of the ion cloud with similar starting positions to pass undisturbed, producing a spatially localized ion packet. This significantly increases the ToF mass resolution and produces well-resolved ion peaks. However, the masses have begun to separate by the time they reach the deflectors, and we probe slightly shifted parts of the ion packets of the two masses. We estimate an uncertainty of 20\% in our measurements due to this effect.

To fit the H$^-$ creation curves we solve the set of equations given by
\begin{align}
	\frac{\mathrm{d}}{\mathrm{d}t}N_\mathrm{H}(t)&= k^\mathrm{gr}_{\mathrm{H}}N_\mathrm{D}(t)-k^\mathrm{bg}_{\mathrm{H}}N_\mathrm{H}(t) \\
	\frac{\mathrm{d}}{\mathrm{d}t}N_\mathrm{D}(t)&= -k^\mathrm{gr}_{\mathrm{H}}N_\mathrm{D}(t)-k^\mathrm{bg}_{\mathrm{D}}N_\mathrm{D}(t),
\end{align}
where $N_\mathrm{X}$ denotes the respective amounts of H$^-$ and D$^-$, $k^\mathrm{bg}_{\mathrm{X}}$ are their respective background loss rates out of the trap, and $k^\mathrm{gr}_{\mathrm{H}}$ is the growth rate of H$^-$. The solution is given by equation (2) in the main text. To achieve a more stable fit a second order Taylor expansion of the solution is used to fit the data at the lowest two densities.

\subsubsection*{Ion Trajectory Simulations and Tsallis distributions}

An ion in a radiofrequency (RF) trap immersed in a cold buffer gas has been found to develop a power-law tail in the velocity distribution \cite{Asvany2009:ijms}. It can be modeled using a non-extensive generalization of entropy proposed by Tsallis \cite{Tsallis1988:jsp} and developed further for a variety of physical systems \cite{Silva1998:pla,Jiulin2004:el,Rouse2017:prl}.

To model the ion distribution we perform molecular dynamics simulations, calculating the ion trajectories by solving Newton's equation of motion numerically, and treating the collisions in a hard-sphere-collision model employing a Monte-Carlo approach. We use an ideal time-varying multipole field in the radial direction and a quadrupole plus octupole static field in the axial direction, along with random collisions with a background buffer gas. Ion-ion interactions are neglected due to the low ion density within the trap. As collision frequency increases, the bulk of the ions move closer to the trap center. Simultaneously the ion number at the outer edges increase (Extended Data Fig.\ 1a). The energies of the outer ions also increase with higher collision frequency (Extended Data Fig.\ 1b). This indicates a runaway heating effect due to multiple sequential heating collisions, which becomes significant when the mean free path of the ions becomes small compared to the mean distance traveled up the trap potential.

We fit each dimension of the velocity distribution of the ions $f_\mathrm{1D}(v_i)$ (see also equation (2)), with $i=x,y,z$, to the $q$-exponential
\begin{equation}
	f_\mathrm{1D}(v_i) = N_\mathrm{1D} \left( 1+(q-1)\frac{m v_i^2}{2k_BT}\right)^{\frac{1}{1-q}},
\end{equation}
where $m$ is the ion mass, $T$ the temperature, and $q$ the parameter giving the strength of the tail, with normalization factor
\begin{equation}
	N_\mathrm{1D} = \left( \frac{m}{2k_BT}\right)^{1/2}\frac{\Gamma\left(\frac{1}{q-1} \right)}{\Gamma\left(\frac{3-q}{2\left( q-1\right)} \right) } \left( \frac{q-1}{\pi}\right)^{1/2}
\end{equation}
for $1<q<3$. The two radial directions exhibit the same distribution due to symmetry. From the one-dimensional velocity distribution one can derive the full distribution of speeds $v$
\begin{equation}
	f_\mathrm{3D}(v) = N_\mathrm{3D} 4\pi v^2\left( 1+(q-1)\frac{m v^2}{2k_BT}\right)^{\frac{1}{1-q}}
\end{equation}
with normalization factor
\begin{equation}
	N_\mathrm{3D} = \left( \frac{m}{2k_BT}\right)^{3/2}\frac{\Gamma\left(\frac{1}{q-1} \right)}{\Gamma\left(\frac{1}{q-1}-\frac{3}{2} \right) } \left( \frac{q-1}{\pi}\right)^{3/2}.
\end{equation}
Equation (8) does not exactly describe the distribution of relative velocities between a Tsallis and a Maxwell-Boltzmann distribution, but can be used empirically as a reasonable fit function (Fig.\ 3c). Since in this work the effects of the high-energy tail are of prime interest, we fit the logarithm of the distribution function (8) to the logarithm of the distribution, which puts more weight on the high energy tail.

The calculations for the reaction rate coefficients without tunneling, shown in Fig.\ 4, were performed using a step function as the energy barrier. The reaction rates come from the percentage of ions that have a collision energy above this barrier. The height of the barrier was adjusted such that the rate without tunneling matches the calculated rate from a Maxwell-Boltzmann distribution at high temperatures from Ref.\ \cite{Yuen2018:pra}. The corresponding barrier height is 275\,meV, which is consistent with the experimentally measured barrier of 330$\pm$60\,meV \cite{Haufler1997:jpca}.



\subsection*{Data availability}

The datasets used for this study are available on zenodo.org at at doi:10.5281/zenodo.7148592.

\subsection*{Code availability}

The code used during this study is available from the corresponding author on reasonable request.

\subsection*{Acknowledgments}

We thank Viatcheslav Kokoouline, Chi Hong Yuen, and Mehdi Ayouz for fruitful discussions. This work has been supported by the Austrian Science Fund (FWF) through Project I2920-N27 and through the Doctoral Programme Atoms, Light, and Molecules, Project No. W1259-N27.

\subsection*{Author contributions}

RWe conceived the experiment and supervized the project. RWi, MN, MS, and DT carried out the measurements. RWi with support from MN and RWe carried out the simulations. RWi and RWe wrote the manuscript, which was discussed and approved by all authors.

\subsection*{Competing interests}

The authors declare no competing interests.

\clearpage

\subsection*{Extended Data}

\vspace{2cm}

\begin{figure}[h]
	\centering
	\includegraphics[width=0.8\columnwidth]{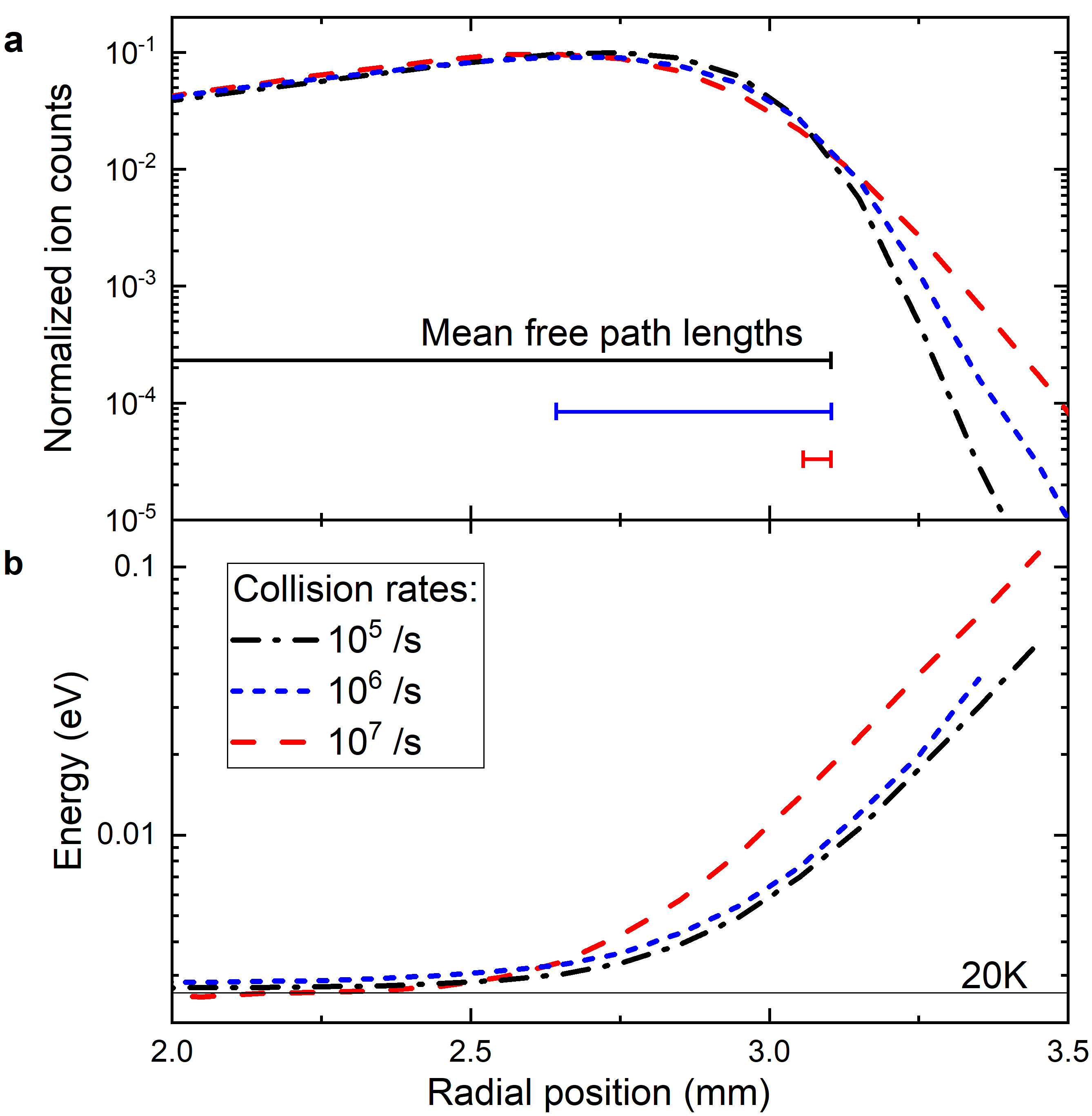}
	\caption{{\bf Ion trap simulations for different densities} Distributions of ion number and energies as a function of the radial position in the trap, at different collision rates and 20\,K buffer gas temperature. {\bf a} With higher collision rates, ions move on average slightly closer to center, but also increase in number at large radii. Mean free path lengths are shown for reference. {\bf b} When the mean free path becomes small, energies close to the trap rods increase significantly, indicating multiple heating collisions before moving away from the trap edges. The collision rates correspond to H$_2$ densities of $4.8\times 10^{13}$, $10^{14}$, and $10^{15}$ cm$^{-3}$, respectively.}
\end{figure}

\begin{figure}[b]
	\centering
	\includegraphics[height=11cm]{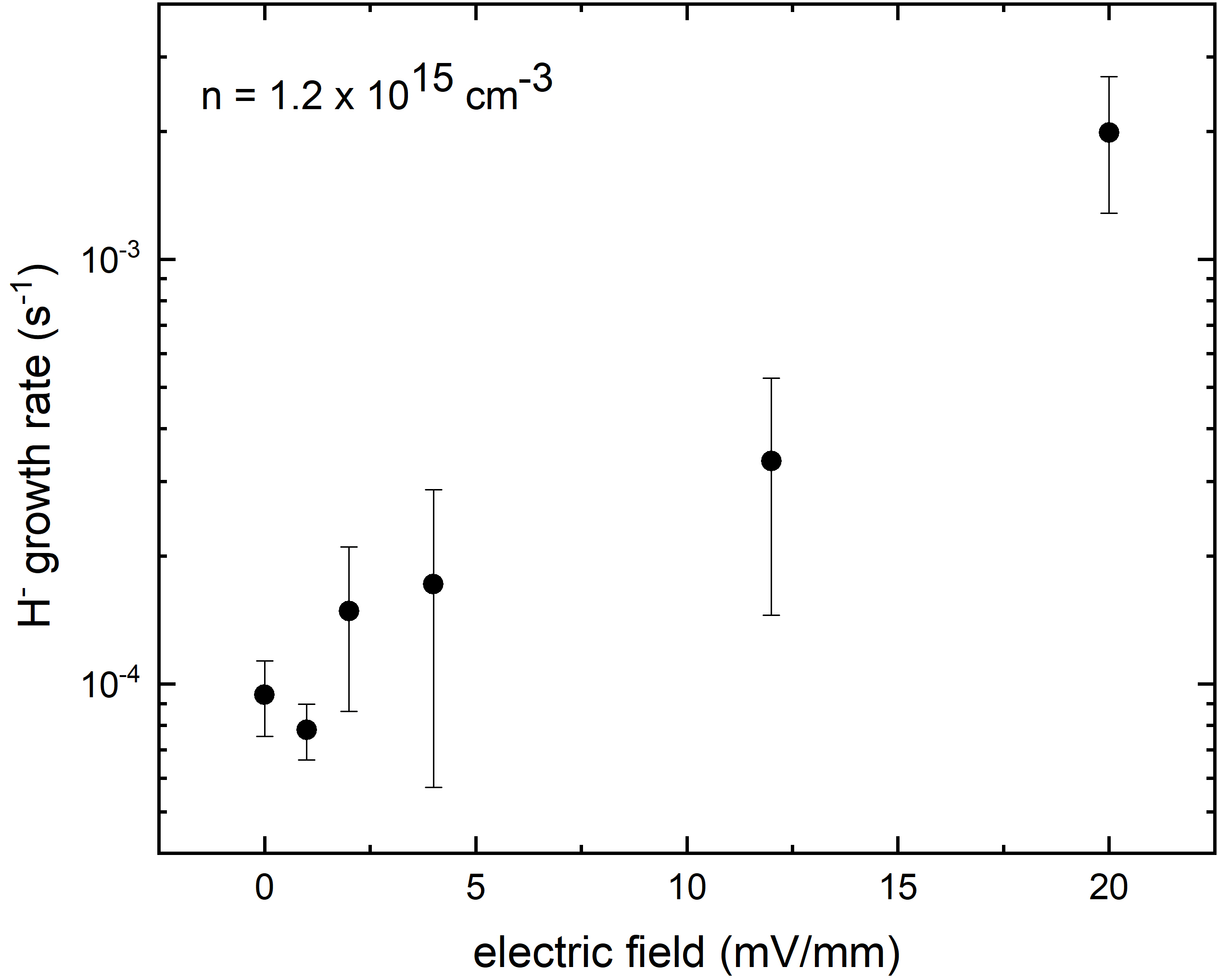}
	\caption{{\bf Ion trap simulations including stray electric fields.} Numerical simulations of the expected reaction rates at a hydrogen gas density of $1.2\times 10^{15}$ $\mathrm{cm}^{-3}$. The electric fields were chosen to be homogeneous and oriented in one direction perpendicular to the rf trapping rods, which has the effect of slightly pushing the ions toward the rf electrodes on one side. The simulations show that an extra field on the order of 10\,mV/mm could account for the higher measured reaction rates at this density.}
\end{figure}

\end{document}